\documentclass[aps,prl,reprint,superscriptaddress]{revtex4-1}

\usepackage{graphicx,hyperref,amsmath,epstopdf,natbib,textcomp,color}

\begin{document}


\title{Electron spin coherence of shallow donors in natural and isotopically enriched germanium}


\author{A. J. Sigillito}
\email[]{asigilli@princeton.edu}
\affiliation{Department of Electrical Engineering, Princeton University, Princeton, New Jersey 08544, USA}

\author{R. M. Jock}
\affiliation{Department of Electrical Engineering, Princeton University, Princeton, New Jersey 08544, USA}

\author{A. M. Tyryshkin}
\affiliation{Department of Electrical Engineering, Princeton University, Princeton, New Jersey 08544, USA}

\author{J. W. Beeman}
\affiliation{Materials Sciences Division, Lawrence Berkeley National Laboratory, Berkeley, California 94720, USA}

\author{E. E. Haller}
\affiliation{Department of Materials Science and Engineering, University of California, Berkeley, California 94720, USA}
\affiliation{Materials Sciences Division, Lawrence Berkeley National Laboratory, Berkeley, California 94720, USA}

\author{K. M. Itoh}
\affiliation{School of Fundamental Science and Technology, Keio University, 3-14-1 Hiyoshi, Kohuku-ku, Yokohama 223-8522, Japan}

\author{S. A. Lyon}
\affiliation{Department of Electrical Engineering, Princeton University, Princeton, New Jersey 08544, USA}


\date{\today}

\begin{abstract}
Germanium is a widely used material for electronic and optoelectronic devices and recently it has become an important material for spintronics and quantum computing applications. Donor spins in silicon have been shown to support very long coherence times ($T_{2}$) when the host material is isotopically enriched to remove any magnetic nuclei. Germanium also has non-magnetic isotopes so it is expected to support long $T_{2}$s while offering some new properties. Compared to Si, Ge has a strong spin-orbit coupling, large electron wavefunction, high mobility, and highly anisotropic conduction band valleys which will all give rise to new physics. In this Letter, the first pulsed electron spin resonance (ESR) measurements of $T_{2}$ and the spin-lattice relaxation ($T_1$) times for $^{75}$As and $^{31}$P donors in natural and isotopically enriched germanium are presented. We compare samples with various levels of isotopic enrichment and find that spectral diffusion due to $^{73}$Ge nuclear spins limits the coherence in samples with significant amounts of $^{73}$Ge. For the most highly enriched samples, we find that $T_1$ limits $T_2$ to $T_2 = 2T_1$. We report an anisotropy in $T_1$ and the ensemble linewidths for magnetic fields oriented along different crystal axes but do not resolve any angular dependence to the spectral-diffusion-limited $T_2$ in samples with $^{73}$Ge.

\end{abstract}


\maketitle



Germanium was the original material for transistors, and is now being developed for the latest semiconductor electronics \cite{lee2005}. Recently, it has become a key material for spintronics \cite{li2013, dushenko2015, shen2010} and quantum computing\cite{vrijen2000, rahman2009, witzel2012} devices. Compared to silicon, donor electrons in Ge have higher mobility ($\sim 3$ times)\cite{lee2005}, larger wavefunctions ($6.5$ nm compared to $2.5$ nm), \cite{wilson1964,feherSi1959}, stronger spin-orbit coupling\cite{liu1962}, and highly anisotropic conduction band valleys \cite{rahman2009}. Much of silicon's success in the quantum computing community has hinged on the attainability of long coherence times ($T_{2}$) exceeding seconds when Si is isotopically enriched to have no magnetic nuclei \cite{morton2011, tyryshkin2012, wolfowicz2013, itoh2014}. Germanium also has non-magnetic isotopes so it has been expected to support long coherence times. In this Letter, we report the first electron spin coherence measurements for donor electron spins in Ge. We find that spectral diffusion due to $^{73}$Ge limits $T_{2}$ in natural Ge samples whereas the spin-lattice relaxation time, $T_{1}$, limits $T_{2}$ in isotopically enriched Ge. The longest $T_{2}$ we measured is $T_{2} = 2T_{1} = 1.2$ ms at 350 mK in a magnetic field ($B_{0}$) of 0.44 T. The low-temperature $T_{1}$ fits the temperature dependence theorized by Roth \cite{roth1960} and Hasegawa \cite{hasegawa1960} which also predicts $T_{1} \propto B_{0}^{-4}$. This suggests that considerably longer coherence times are possible at lower fields.

While $T_{2}$ for donors in Ge is shorter than the times demonstrated for Si, Ge-based qubits have some important advantages. For example the larger electron wavefunctions relax the lithographic requirements for exchange coupling two donors which is important for most donor-based quantum computing schemes \cite{kane1998}. This is advantageous considering Ge is compatible with most of the same nanofabrication techniques as silicon and single-donor devices are achievable \cite{scappucci2011}. Another useful feature of Ge is the large spin-orbit coupling and shallow donor depth which leads to a very large spin-orbit Stark shift in Ge (nearly 5 orders of magnitude larger than in silicon) \cite{rahman2009} meaning that Ge based qubits are extremely tuneable. This will be important for gated quantum devices\cite{kane1998}.

Despite these features, the spin coherence of donor electrons in Ge has remained mostly unstudied. The first experiments were conducted over fifty years ago by Feher, Wilson, and Gere \cite{Feher1959, wilson1964}, but their measurements were limited to continuous wave (CW) ESR spectroscopy. They estimated $T_{1}$ for $^{75}$As and $^{31}$P donors based on power saturation measurements, but experimental errors were large. These experiments are difficult because wavefunction overlap occurs for densities as low as $10^{15}$ donors/cm$^{3}$ such that only lightly doped samples with correspondingly weak signals are useful for isolated donor experiments. Some limited experiments on Sb \cite{Pontinen1966, hale1975} and $^{31}$P \cite{morigaki1963} donors in highly strained Ge were also reported. More recently, pulsed nuclear magnetic resonance studies were conducted on $^{73}$Ge nuclear spins \cite{verkhovskii1999, verkhovskii2003, panich2007} which found that the $^{73}$Ge nuclear spin coherence in germanium can be $>100 \ ms$.

The samples discussed in this Letter include commercially available, natural Ge doped either $10^{15}$ As/cm$^{3}$ or $10^{12}$ P/cm$^{3}$. $^{73}$Ge is the only naturally occurring isotope of Ge (7.75\% abundance) with a nuclear spin and is thus expected to be a limiting factor in the donor spin coherence at low temperatures. Three isotopically enriched samples were prepared at Lawrence Berkeley National Laboratory. The first is a piece of neutron transmutation doped $^{74}$Ge described in Ref.\cite{itoh1993, itoh1994}. This sample is uniformly doped with $^{75}$As to a density of $3 \times 10^{15}$ donors/cm$^3$ and contains a residual 3.8\% $^{73}$Ge. The other two samples are 96\% $^{70}$Ge crystal (0.1\% $^{73}$Ge) and a 99.99\% $^{70}$Ge crystal (0.01\% $^{73}$Ge). They have $^{31}$P concentrations of $\sim 10^{12}$ donors/cm$^3$ and $\sim 10^{11}$ donors/cm$^{3}$, respectively and are described in \cite{itoh1993,Palmer1997}. The crystallographic orientation of the samples was determined using X-ray diffraction. The sample details are summarized in Table~\ref{table:table1}.

\begin{table*}
\caption{Sample Details} 
\centering 
\begin{tabular}{|c| c c c c c| c | c | c|}
\hline\hline 
Sample Name  &  $[^{70}$Ge$]$  & $[^{72}$Ge$]$ & $[^{73}$Ge$]$  &  $[^{74}$Ge$]$  & $[^{76}$Ge$]$ & Doping (cm$^{-3})$ & [001] Linewidth (mT) & $T_2^*$ (ns) \\ [0.5ex]
\hline 
$^{nat}$Ge:As$^{*}$ & 20.57\% & 27.45\%    & 7.75\%   & 36.50\% & 7.73\% & $1 \times 10^{15}$ As & 1.2 & 11  \\ 
$^{nat}$Ge:P$^{*}$ & 20.57\%    & 27.45\%   & 7.75\%   & 36.50\% & 7.73\% & $\sim 10^{12}$ P & 1.1 & 13 \\
3.8\%$^{73}$Ge:As & 0.1\% & 0.9\%   & 3.8\%   & 92.6\% & 2.6\% & $3 \times 10^{15}$ As & 0.8 & 17\\
0.1\% $^{73}$Ge:P & 96.3\%    & 2.1\% & 0.1\% & 1.2\% & 0.3\% & $\sim 10^{12}$ P & 0.069 & 211\\
0.01\% $^{73}$Ge:P & 99.99\%    &  -    & 0.01\%  & - & -  & $\sim 10^{11}$ P & 0.051 & 284\\ 
\hline
Nuclear Spin & 0 & 0 & 9/2 & 0 & 0 & - & - & - \\
\hline 
\end{tabular}
\label{table:table1} 

$^*$Percent abundances for the natural germanium samples were taken from Ref. \cite{berglund2009}
\end{table*}

\begin{figure} [h]
\includegraphics{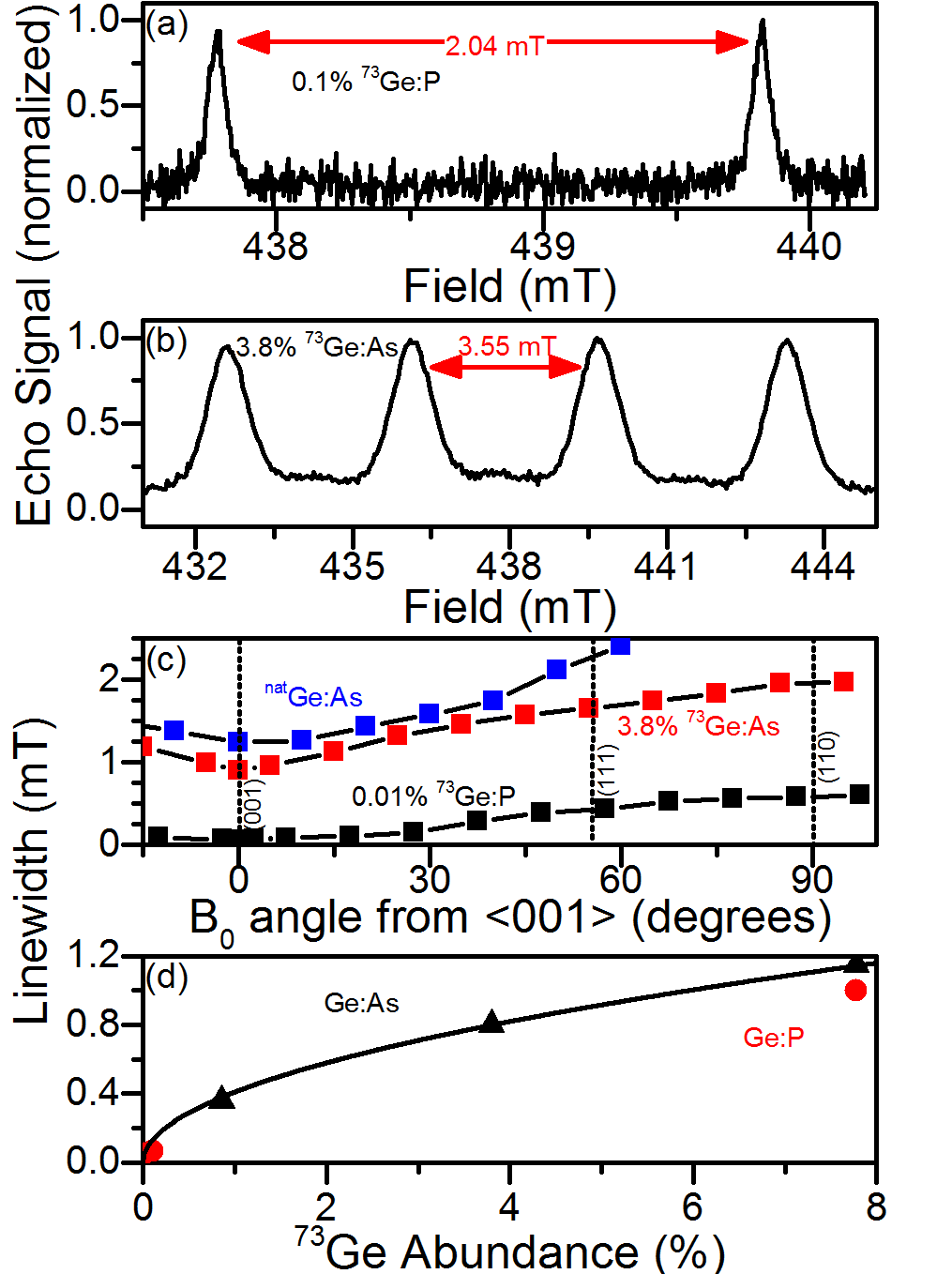}
\caption{\label{fig:fig1} (a) Echo-detected field sweep spectra for (a) 0.1\% $^{73}$Ge:P and (b) 3.8\% $^{73}$Ge:As with $B_0 \ || \ \langle 001 \rangle$. (c) Plot of ESR linewidths as a function of field orientation for $^{nat}$Ge:As (blue), 3.8\% $^{73}$Ge:As (red) and 0.01\% $^{73}$Ge:P(black). The solid lines serve only as guides to the eye. (d) Linewidth for $B_0 \ || \ \langle 001 \rangle$ as a function of $^{73}$Ge isotopic abundance. The Ge:As data appear as black triangles whereas the Ge:P data appear as red circles. The solid line shows the expected $f^{1/2}$ dependence for broadening due to $^{73}$Ge hyperfine interactions. The ESR linewidth at 0.8\% is taken from Ref. \cite{wilson1964}. Data were taken at 1.8 K and 9.65 GHz.} 
\end{figure}

The experiments down to 1.65 K were performed in a pumped He cryostat (H.S. Martin), and lower temperature data were obtained in a $^3$He cryostat (Janis Research). All data were taken at X-band (9.65 GHz) in a Bruker dielectric resonator (MD5). The ESR spectra were measured via echo-detected field sweeps using a standard Hahn-echo pulse sequence ($\pi$/2 - $\tau$ - $\pi$ - $\tau$ - echo). Typical spectra are shown in Fig.~\ref{fig:fig1}(a) for phosphorus donors in the 0.1\% $^{73}$Ge:P sample and in Fig.~\ref{fig:fig1}(b) for arsenic donors in the 3.8\% $^{73}$Ge:As sample. From these plots we extract a hyperfine coupling constant of 3.55 mT for $^{75}$As and 2.04 mT for $^{31}$P.

The ESR linewidth depends strongly on the sample orientation and the abundance of $^{73}$Ge present in the sample, as noted by Wilson \cite{wilson1964}. With $B_0$ oriented along one of the $\langle 001 \rangle$ directions, the linewidth is narrowest and is limited primarily by hyperfine interactions with $^{73}$Ge. At this orientation the line broadening from spin-orbit strain effects is suppressed by valley symmetry about the $\langle 001 \rangle $ as explained in Refs.\cite{wilson1964, roth1960, hasegawa1960}).  To give a sense of the strain-induced line broadening for $B_{0}$ away from $[$001$]$ equivalent directions, Fig.\ref{fig:fig1}(c) shows the angular dependence of the linewidth for select samples rotated in the (1$\overline{1}$0) plane relative to the $[001]$ axis. There is also an isotopic dependence of the linewidth away from the $\langle 001 \rangle$ direction and we presume this is due to isotopic strain \cite{Stegner2010}. The strong dependence of the linewidth on field orientation conveniently allows for accurate \textit{in situ} orientation of the crystals. Unless otherwise noted, all data presented in this manuscript assumes $B_{0}$ is oriented along a $\langle 001 \rangle$ axis.

One can predict the effect of $^{73}$Ge on the ESR linewidth through the hyperfine interaction using a second moment calculation \cite{Kittel1953}, which gives $\Delta B \propto f^{1/2}$, where $\Delta B$ is the linewidth, and $f$ is the percent abundance of $^{73}$Ge. The measured ESR linewidths for samples of various isotopic enrichment with $B_{0} || \langle 001 \rangle$ is shown in Fig.~\ref{fig:fig1}(d). The point at $f$ = 0.8\% was taken from Wilson \cite{wilson1964}. The solid curve in Fig. \ref{fig:fig1}(d) gives the expected $f^{1/2}$ dependence for broadening of the line due to $^{73}$Ge hyperfine interactions for $^{75}$As. The solid curve fits the data well, implying that $^{73}$Ge is indeed the dominant mechanism for line broadening in this orientation. The linewidth can be interpreted as an ensemble dephasing time, $T_{2}^{*}$, which is also shown in Table~\ref{table:table1}.

\begin{figure} [h]
\includegraphics{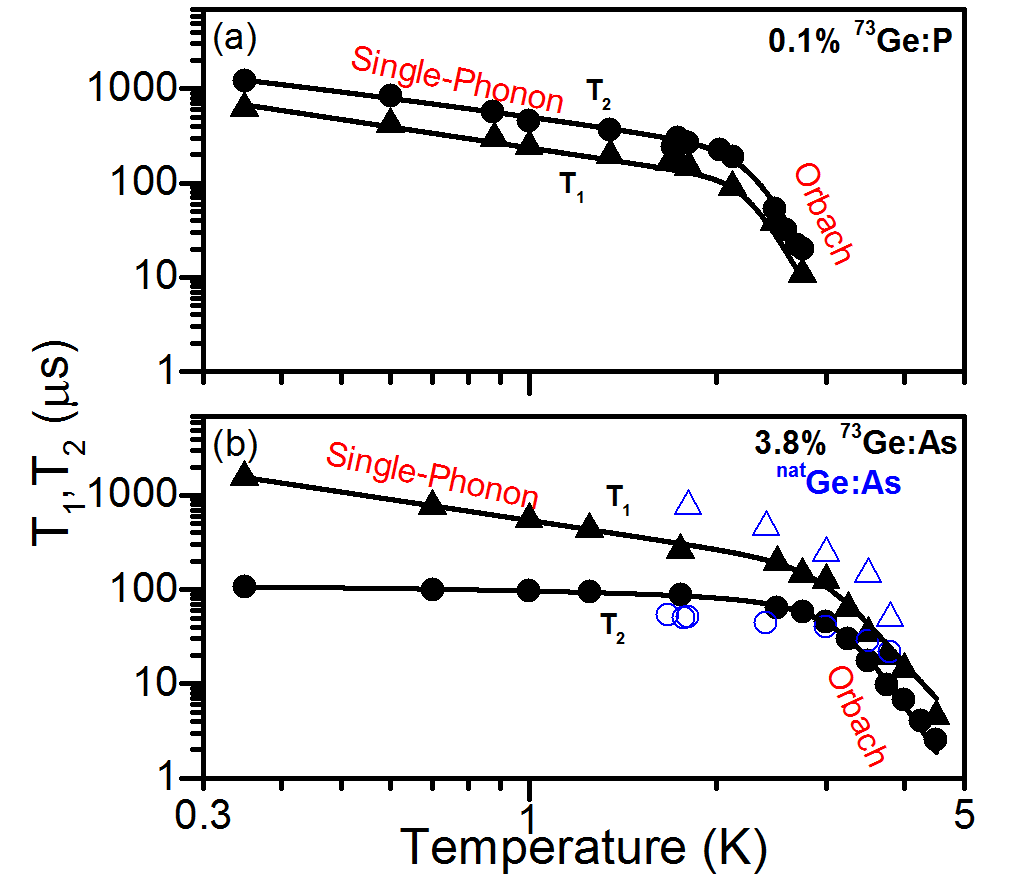}
\caption{\label{fig:fig2} Temperature dependence of $T_{1}$ (triangle) and $T_{2}$ (circle) for natural (open symbols) and isotopically-enriched (solid symbols) Ge with $B_0 \ || \ \langle 001 \rangle$. The solid lines are fits for (a) phosphorus donors (0.1\% $^{73}$Ge:P)  and (b) arsenic donors (3.8\% $^{73}$Ge:As), assuming two relaxation processes: a single-phonon ($T^{-1}$) process and an Orbach ($a \times exp(\frac{E_{v.o.}}{kT})$) process. For the $T_{2}$ fits, both $T_{1}$ and an additional (temperature independent) spectral diffusion mechanism due to $^{73}$Ge were taken into account. Note that for the 0.1\% $^{73}$Ge:P sample, $T_2 = 2T_1$ down to the lowest temperatures. } 
\end{figure}

$T_{1}$ was measured using an inversion-recovery pulse sequence ($\pi$ - t - $\pi$/2 - $\tau$ - $\pi$ - $\tau$ - echo). The values of $T_{1}$ are plotted in Fig.~\ref{fig:fig2} for $^{31}$P(a) and $^{75}$As(b) donors. The same two mechanisms limit $T_{1}$ for all of the samples. At higher temperatures, $T_{1}$ is limited by a highly temperature ($T$) dependent process. The theory of Roth and Hasegawa \cite{roth1960, hasegawa1960} predicted a $T^{-7}$ Raman process to dominate at these temperatures but this dependence does not fit our data well. An Orbach process does fit the data as shown in Fig.~\ref{fig:fig2}. The Orbach process is of the form $T_{1} \propto a \times exp(\frac{E_{v.o.}}{kT})$, where $a$ is a prefactor that can be calculated using Ref. \cite{Castner1967}, $E_{v.o.}$ is the valley-orbit splitting, and $k$ is the Boltzmann constant. The valley-orbit splittings extracted from the $T_{1}$ fits in Fig.~\ref{fig:fig2} agree well with the values measured by Ramdas ($2.8$ meV for $^{31}$P and $4.2$ meV for $^{75}$As \cite{ramdas1981}). Likewise, the values of $a$ extracted from our fits agree with the values calculated using Castner's theory \cite{Castner1967} to within a factor of 2.

At lower temperatures, a single-phonon process with a $T^{-1}$ dependence appears to dominate. This relaxation process is a result of the multivalley structure of germanium. In the unperturbed ground state, there are four degenerate valleys located along the $\langle 111 \rangle$ equivalent crystallographic axes. Each valley has an axially symmetric g-tensor, $\tensor{g}_i$ but the effective g-tensor, $\tensor{g}_{eff}$, is given as a weighted average over all four valleys. In the electron ground state, each valley has equal amplitude, and, by symmetry, $\tensor{g}_{eff}$ is isotropic \cite{roth1960}. When strain is applied, valley energies shift relative to each other, leading to valley repopulation and a change in $\tensor{g}_{eff}$. The strain from phonons near the Larmor frequency modulates $\tensor{g}_{eff}$, effectively mixing the spin up and down states. This gives a $T_{1}$ as calculated by Roth \cite{roth1960} and Hasegawa \cite{hasegawa1960} which agrees well with our experimental data. The calculated estimates for $T_{1}$ at 350 mK are within 10\% for Ge:As and 30\% for Ge:P. The theory predicts that $T_1$ due to this single-phonon process should scale with the square of the $\tensor{g}_i$ anisotropy. The valley anisotropy of Ge was measured to be 3 orders of magnitude larger than in Si \cite{wilson1964}, implying that the single-phonon process should be 6 orders of magnitude stronger in germanium. This accounts for the short $T_{1}$ times observed for donors in germanium as compared with silicon.

\begin{figure} [h]
\includegraphics{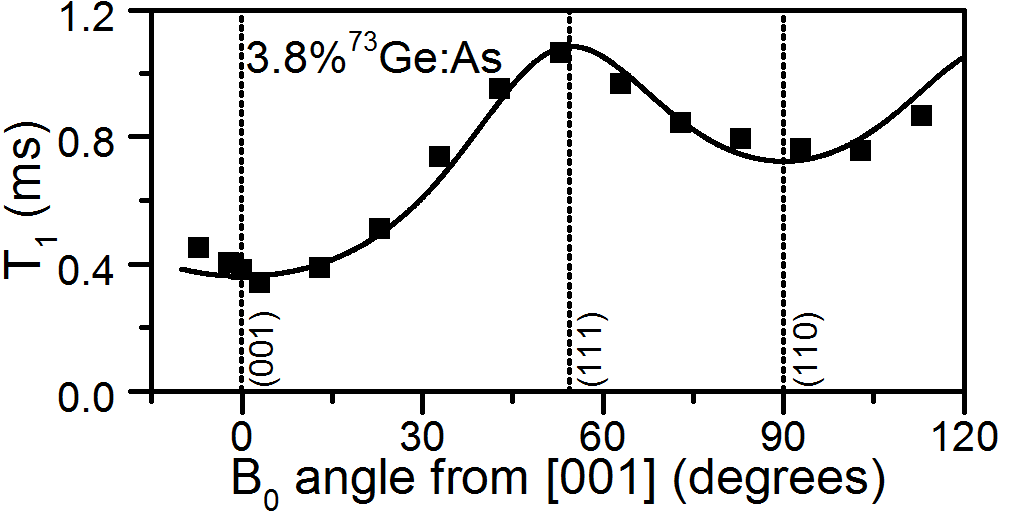}
\caption{\label{fig:fig3} Angular dependence of $T_{1}$ for the 3.8\% $^{73}$Ge:As sample rotated in the $(1\overline{1}0)$ plane at 1.8 K. The curve is a fit using Eq.(\ref{equation:eq1}) assuming $\alpha = 4.1 \times 10^{4} \ K^{-1}s^{-1}T^{-4}$.}
\end{figure}

An interesting property of the single-phonon spin-lattice relaxation mechanism is an anisotropy in $T_1$ predicted by the Roth-Hasegawa theory\cite{roth1960, hasegawa1960}. The 3.8\% $^{73}$Ge:As crystal was rotated in the $(1\overline{1}0)$ plane at 1.8 K, and the resulting $T_{1}$ is plotted in Fig.\ref{fig:fig3}. The theory predicts that, for rotation in this plane, the spin-lattice relaxation is given by:
\begin{equation}\label{equation:eq1}
\frac{1}{T_{1}} = \alpha B_{0}^4 T (cos^{4}(\theta) + \frac{1}{2} sin^{4}(\theta))
\end{equation}
where $\alpha$ is a scaling factor which can be calculated following Hasegawa\cite{hasegawa1960}, and $\theta$ is the field orientation relative to $\langle 001 \rangle$. Hasegawa calculated $\alpha = 7.2 \times 10^{4} K^{-1}s^{-1}T^{-4}$ for arsenic in Ge, but a fit to the data reveals $\alpha = 4.1 \times 10^{4} K^{-1}s^{-1}T^{-4}$. We observe that for $B_0$ oriented along a $\langle 111 \rangle$ axis, $T_1$ becomes 3 times longer than along $\langle 001 \rangle$.

We note that $T_{1}$ for donors in highly enriched samples is shorter than it is for donors in the natural material as seen in Fig.~\ref{fig:fig2}(b). This effect is still under investigation, but one possible mechanism is the presence of isotopic strain in the natural germanium\cite{Stegner2010}. Wilson \cite{wilson1964} demonstrated the use of large strains to partially lift the valley degeneracy, thus disrupting the single-phonon relaxation mechanism. Modelling the effects of strain can be complex, as strain not only modulates $\alpha$, but can also modify the form of Eq.(\ref{equation:eq1}). Nevertheless, controlled strain may be beneficial for future quantum devices based on germanium.

\begin{figure}
\includegraphics{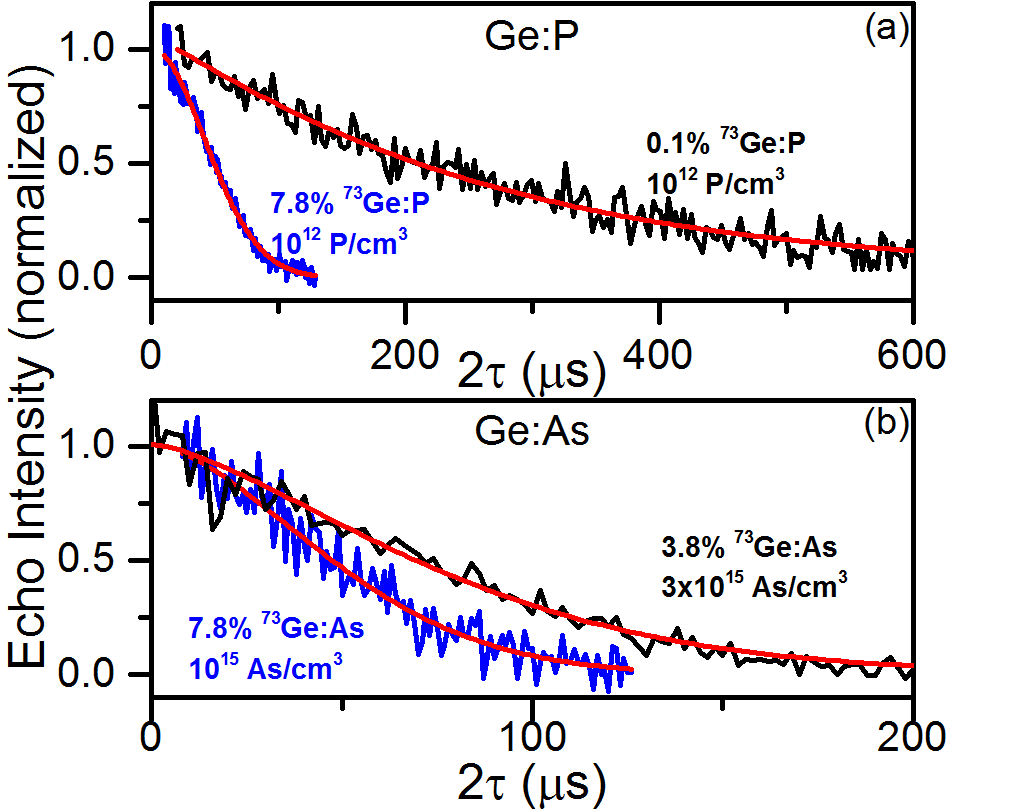}
\caption{\label{fig:fig4} Two-pulse Hahn echo decay curves for natural (blue) and isotopically enriched (black) germanium doped with phosphorus(a) and arsenic(b) donors. Data were taken at 1.8 K and 9.65 GHz. The solid curves are fits to the data using $exp[-(\frac{2\tau}{T_{2}})^{n}]$.} 
\end{figure}

We also measured the electron spin coherence time, $T_{2}$, for each of the samples using the standard Hahn-echo pulse sequence. The decay curves at 1.8 K for $B_{0} \ || \ \langle 001 \rangle$ are shown in Fig.~\ref{fig:fig4}(a) for Ge:P and in Fig.~\ref{fig:fig4}(b) for Ge:As. These decays are fit to an exponential decay of the form $Ae^{-(2\tau/T_{2})^n}$, where $A$ scales the amplitude, $\tau$ is the delay between the $\pi/2$ and $\pi$ pulses in the Hahn echo sequence, and $n$ is a fitting parameter that depends on the decoherence mechanism. The 0.1\% $^{73}$Ge:P sample decays with $n$=1 over the measured temperature range. For this sample it was found that $T_{2} = 2T_{1}$ (representing the absolute $T_{1}$ limit \cite{Schweiger2001}) down to 350 mK temperatures, meaning that decoherence due to $^{73}$Ge is negligibly small with this level of isotopic enrichment at these temperatures. For samples with $f \geq 3.8\%$, we find that $n$ varies from 1 at high temperatures to 2.1 at low temperatures. This is a characteristic of $^{73}$Ge spectral diffusion limiting the coherence. At 1.8 K, the $^{nat}$Ge:As, $^{nat}$Ge:P, and 3.8\% $^{73}$Ge:As samples decay with this form. 

The temperature dependence of $T_{2}$ is also plotted in Fig.~\ref{fig:fig2} and fit to $1/T_{2} = 1/(2T_{1}) + 1/T_{SD}$, where $T_{SD}$ is the (temperature independent) spectral-diffusion-limited coherence time. For the natural germanium samples, $T_{SD}$ limits the coherence to 57 $\mu$s whereas the 3.8\% $^{73}$Ge:As sample is limited to 113 $\mu$s. From similar work in silicon \cite{abe2010, desousa2003}, one might expect an orientation dependence to $T_{SD}$. We measured the orientation dependence of $T_{SD}$ for the 3.8\% $^{73}$Ge:As sample at 1.8 K and fit the decays with a curve of the form  $Ae^{-(2\tau/(2 T_{1})}e^{-(2\tau/T_{SD})^n}$ to separate the $T_1$ component from $T_{SD}$ \cite{mims1961}.  No angular dependence of $T_{SD}$ could be resolved.

While coherence times of over one millisecond for isotopically enriched material open the possibility of using donor electrons in Ge for quantum computing devices, these coherence times are much shorter than those for donors in isotopically enriched silicon (seconds) \cite{tyryshkin2012, wolfowicz2013}. To extend the Ge donor coherence, one must either overcome the $T_{1}$ limit or use nuclear spins which may support longer coherence times. There are several promising techniques to extend the $T_{1}$ limit. One approach is to take advantage of the $T_1$ anisotropy, which will allow for up to a factor of 3 increase in $T_1$ when devices are oriented with $B_0 \ || \ \langle 111 \rangle$, but this $T_1$ enhancement comes at the expense of a shorter ensemble $T_2^*$. A simple alternative is to operate devices at lower temperatures, since $T_{1} \propto T^{-1}$. Perhaps the most effective technique is to operate devices at lower frequencies since theory predicts $T_{1} \propto B_{0}^{-4}$. More complicated strategies are also available. In particular, one can apply a large strain, as demonstrated by Wilson \cite{wilson1964} which shifts the valley energy levels, thus suppressing valley repopulation and the associated relaxation mechanisms. Another recent proposal suggests patterning Ge in a periodic structure to open a phononic bandgap at the Larmor frequency\cite{Smelyanskiy2014}. Such a structure would suppress the single phonon process.

In summary, we have measured the ESR linewidths, coherence times, and spin-lattice relaxation times for donors in natural and isotopically enriched germanium at X-band microwave frequencies. We find that the linewidths are primarily broadened by hyperfine interactions with $^{73}$Ge spins when $B_0$ is oriented along the $[$001$]$ axis and by strain in other orientations. We find that donor electron spin coherence is limited by spectral diffusion due to hyperfine interactions with $^{73}$Ge nuclei for the $^{nat}$Ge ($T_{SD} = 57 \ \mu s$) and 3.8\% $^{73}$Ge:As ($T_{SD} = 113 \ \mu s$) samples, thus $T_{SD}$ scales approximately as $1/f$ which is similar to silicon\cite{desousa2003}. For the more highly enriched 0.1\% $^{73}$Ge:P sample, $T_{2}$ was limited to $2 T_{1}$ down to 350 mK, the lowest temperature we have measured \ ($T_{2} = 1.2$ ms for  $B_0 \ || \ \langle 001 \rangle$). We observe a large anisotropy in $T_1$, which is explained by the theory of Roth and Hasegawa\cite{roth1960, hasegawa1960}, with the longest $T_1$ occuring for $B_0 \ || \ \langle 111 \rangle$. It is predicted that at lower magnetic fields $T_1$ and thus $T_2$ should become substantially longer.


\subsection{}
\subsubsection{}

\begin{acknowledgments}
Work at Princeton was supported by the NSF through the Materials World Network and MRSEC Programs (Grant No. DMR-1107606 and DMR-01420541), and the ARO (Grant No. W911NF-13-1-0179). The work at Keio has been supported by the Core-to-Core Program by JSPS, and the Grants-in-Aid for Scientific Research, and Project for Developing Innovation Systems by MEXT.
\end{acknowledgments}

\providecommand{\noopsort}[1]{}\providecommand{\singleletter}[1]{#1}%

\end{document}